\documentclass[draft]{mn2e}
\usepackage{graphicx}
\usepackage{epstopdf}
\usepackage{subfigure}
\usepackage{lscape}
\usepackage{longtable}
\usepackage{wasysym}
\usepackage{psfig}

\usepackage{wasysym}

\def\gs{\mathrel{\raise0.35ex\hbox{$\scriptstyle >$}\kern-0.6em
\lower0.40ex\hbox{{$\scriptstyle \sim$}}}}
\def\ls{\mathrel{\raise0.35ex\hbox{$\scriptstyle <$}\kern-0.6em
\lower0.40ex\hbox{{$\scriptstyle \sim$}}}}
\def\ls{\mathrel{\hbox{\rlap{\hbox{\lower4pt\hbox{$\sim$}}}\hbox{$<$}}}}
\def\gs{\mathrel{\hbox{\rlap{\hbox{\lower4pt\hbox{$\sim$}}}\hbox{$>$}}}}

\def\mnras {{\sc MNRAS}}

\title[Antlia Dwarf Galaxy]
      {Antlia Dwarf Galaxy: Distance, quantitative morphology and recent formation history via statistical field correction}
\author[K.\ A.\ Pimbblet and W.\ J.\ Couch]
       {Kevin A.\ Pimbblet\,$^{1,2}$\thanks{email: Kevin.Pimbblet@monash.edu}, 
	Warrick J.\ Couch\,$^{3}$
        \vspace*{1mm}\\
        $^{1}$School of Physics, Monash University, Clayton, Victoria 3800, Australia\\
$^{2}$Monash Centre for Astrophysics (MoCA), Monash University, Clayton, Victoria 3800, Australia\\
$^{3}$Centre for Astrophysics and Supercomputing, Swinburne University of Technology,
Hawthorn, Victoria 3122, Australia\\
}

\date{\fbox{\sc Draft: \today\ --- Do Not Distribute}}

\pagerange{000--000}

\begin{document}

\maketitle

\begin{abstract}
We apply a statistical field correction technique originally 
designed to determine membership of high redshift galaxy clusters to Hubble Space
Telescope imaging of the Antlia Dwarf Galaxy; 
a galaxy at the very edge of the Local Group.  
Using the tip of the red giant branch standard candle method
coupled with a simple Sobel edge detection filter
we find a new distance to Antlia of $1.31\pm0.03$ Mpc.
For the first time for a Local Group Member, we
compute the concentration, asymmetry and clumpiness (CAS) 
quantitative morphology parameters 
for Antlia from 
the distribution of resolved stars in the HST/ACS 
field, corrected with a new method for contaminants and
complement these parameters with the Gini coefficient ($G$) and 
the second order moment of the brightest 20 per cent of the flux ($M_{20}$).
We show that it is a classic dwarf elliptical
($C = 2.0$, $A = 0.063$, $S = 0.077$, $G = 0.39$ and
$M_{20} = -1.17 $ in the F814W band), 
but has an appreciable blue stellar population 
at its core, confirming on-going star-formation.  
The values of asymmetry and clumpiness, as well as Gini and
$M_{20}$
are consistent with an undisturbed galaxy. 
Although our analysis
suggests that Antlia may not be tidally influenced
by NGC~3109 it does not necessarily
preclude such interaction.

\end{abstract}


\section{Introduction}

In the successful hierarchical cold dark matter 
paradigm, galaxies grow through repeated mergers
with other galaxies (e.g.\ Guo \& White 2008; 
De Lucia \& Blaizot 2007; White \& Rees 1978).  
Inside this hierarchy, the dwarf galaxy sits at
the bottom; analogous to a fundamental galaxy 
`building block' that can be combined with 
other blocks in a large variety of ways 
(cf.\ Durhuus \& Eilers 2005).
Indeed, in the local Universe, both the Milky Way and
Andromeda are observed to still be under-going
construction due to the accretion of such
dwarf galaxies (Martin et al.\ 2004; Ibata et al.\ 2001).
Further, the Milky Way, Andromeda and M33 seem to be
the few galaxies in the Local Group that are
not dwarfs -- Mateo (1998) reports that there
are likely in excess of 40 bona fide dwarf
galaxy members of the Local Group (see also Grebel 1997).

Dwarf galaxies are not only the basic building
block for galaxy evolution, but they are
also the most numerous 
across all redshifts (Marzke \& Da Costa 1997; 
Ferguson \& Binggelli 1994).
The Local Group presents a solid test bed for
studying the varied properties of dwarf galaxies.
Often, they appear to have had strong 
(sometimes on-going) star-formation
whose origin is somewhat enigmatic (Mateo 1998 and 
references therein) but
probably triggered by recent (tidal) 
interactions with their close neighbours (see Tolstoy,
Hill \& Tosi 2009 for a detailed review; Lewis et al. 2007).  
Mateo (1998) further point out that it is the 
case that no two Local Group dwarfs have the 
exact same star-formation history (see also
Koleva et al.\ 2009; Weisz et al.\ 2011).


In the present work, we focus on the Antlia Dwarf Galaxy
with the broad aims of discerning its distance, morphology
and recent star-formation history through the use
of an extensively used extra-galactic
contamination subtraction technique (Pimbblet et al.\ 2002)
and a quantitative morphology approach (Conselice 2003).

Although Antlia was noted by Corwin, 
de Vaucouleurs \& de Vaucouleurs (1985) 
as a possible local dwarf galaxy, 
it was Whiting, Irwin \& Hau (1997) who published its distance 
for the first time and
confirmed it as being a probable member of the Local Group.  
Whiting et al.\ (1997) 
suggest that Antlia is a `typical' dwarf elliptical
galaxy, reminiscent of the Tucana dwarf and the various
Milky Way satellites.  
More recent publications suggest that Antlia is anything
but a regular dwarf elliptical, with a strong blue
stellar component and on-going star-formation 
(e.g.\ Aparicio et al.\ 1997; 
Sarajedini et al.\ 1997; Piersimoni et al.\ 1999;
Dalcanton et al.\ 2009; McQuinn et al.\ 2010).
McQuinn et al.\ (2010) made a study of the star-formation
histories of 18 dwarf galaxies that appear to be under-going
star-bursts.  Taking data from the Hubble Space Telescope (HST)
archive, they suggest that the majority of their sample
is still under-going present-day starbursts, whilst
$\sim$30 per cent have indicators of `fossil' star-bursting
within the past few hundred Myr or so.
Amongst those with a
fossil burst, McQuinn et al.\ (2010) note that the Antlia
Dwarf Galaxy has both the lowest mass and star-formation   
rate in their sample.  However, set against the context
of its own history, the fossil burst in Antlia is both
significant and observationally measurable.  Yet, this
galaxy would not be considered to have a significant
star-formation rate from a simple analysis
of its archival ground-based imaging.

Antlia presents an unique target since it is located
on the edge of the Local Group and may have had relatively
few interactions with other group members; its nearest
neighbour being NGC~3109.  Given the distance between
these two galaxies may be as large as 180 kpc
and their relative velocity 45 km s$^{-1}$
(Aparicio et al.\ 1997), it is unlikely they
are gravitationally bound and interacting at present.
But if the distance difference 
is much lower, (e.g.\ $\sim$28 kpc due to them being 
at the same radial distance;
Aparicio et al.\ 1997), then it may be the case
that Antlia is a satellite of NGC~3109 (as suggested by
van den Bergh 1999)
and have had historic interactions with it.
Newer measurements of the distance to Antlia 
(Dalcanton et al.\ 2009) suggest that it could be much
further away -- perhaps over 300 kpc.
Yet, a number of authors suggest that warping in the
disk of NGC~3109 may be due to interaction with Antlia
(Lee, Grebel \& Hodge 2003;
Grebel, Gallagher \& Harbeck 2003;
Barnes \& de Blok 2001;
Jobin \& Carignan 1990).  Our approach to
determining Antlia's quantitative morphology 
will help address both its distance and recent
evolution.

The format of this work is as follows.
In Section 2, we detail the HST dataset that is used in
this work and indicate how we sample the stellar contaminants.
In Section 3, we fully detail the contamination correction
technique of Pimbblet et al.\ (2002) and how we modify it 
to be better suited to the present case.
To calculate the distance to Antlia, we employ a
tip of the red giant branch standard candle method 
in Section 4.  Section 5 details our investigation
of the morphology of Antlia 
using the Conselice (2003) CAS parameters, as well
as the Gini and $M_{20}$ parameters,
and we summarize our findings in Section 6.

\section{Data}
In this work, we utilize the archival HST F814W and F606W
passband observations (i.e.\ a single colour) 
that have been processed by 
the Advanced Camera for Surveys (ACS) Nearby Galaxy
Treasury Survey (ANGST) of Dalcanton et al.\ (2009)
from an original survey by Tully et al.\ (2006).
Here, we summarize the pertinent points from Dalcanton et 
al.\ (2009) concerning the ANGST data pipeline and refer
the reader to that article for a full description and 
treatment.
ANGST is more than an HST imaging survey: 
systematic object detection,
classification and quality control checks have been implemented
uniformly across legacy images of nearby galaxies.

Following standard image reduction steps (flat fielding, bias subtraction),
photometry is performed with the {\sc dolphot} 
package (Dolphin 2000).  The package aligns the stars contained
in the ACS images to a high precision ($\sim0.01$ arcsec) 
and calculates a local point spread function 
for each star from Tiny Tim (Krist 1995).  The flux (and hence magnitude)
for each star is then determined in an iterative process by calculating 
the flux arising from each star in the crowded field.  The final
catalogue is then culled of objects of low `sharpness' (i.e.\ probable 
galaxies) and stars from highly crowded regions whose photometry
may be significantly compromised.

\begin{figure*}
\centerline{\psfig{file=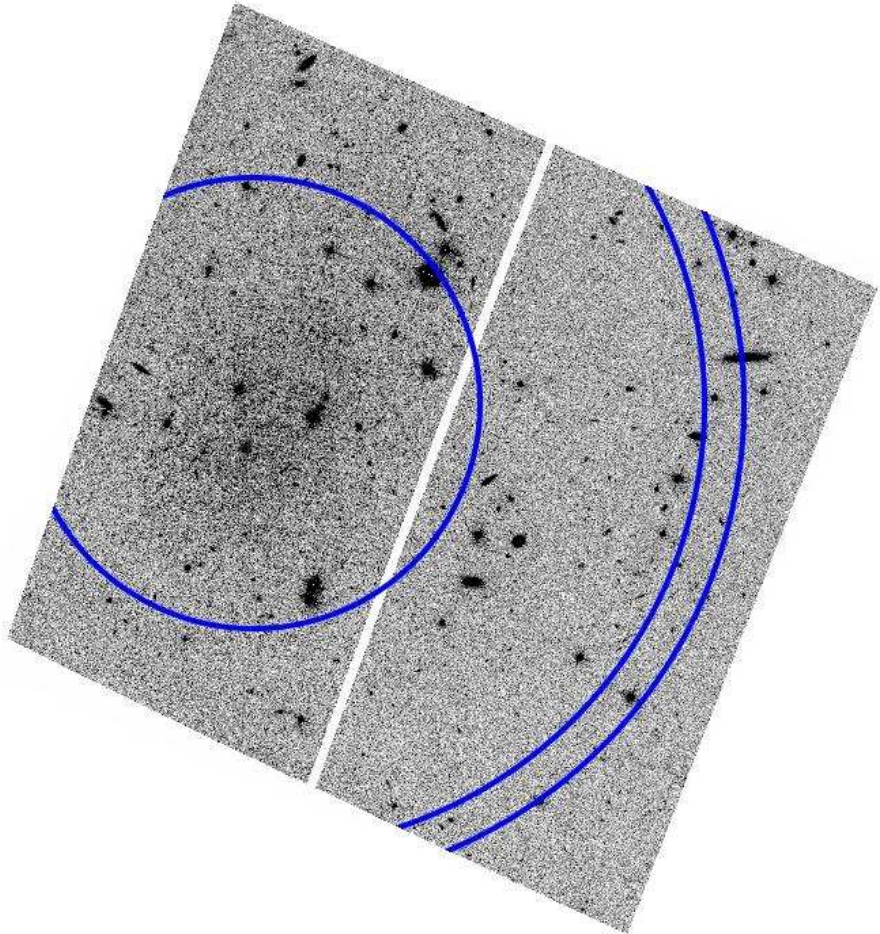,angle=0,width=5in}}
  \caption{HST ACS F814W image of Antlia with
schematic of the divisions between the different samples 
(North is up; East is left).  
Stars contained inside
the inner circle belong to the `inner' sample and are representative
of the high stellar density core of the Antlia dwarf galaxy.  
Stars outside this radius, but within the second concentric circle
belong to the `outer' sample.  Beyond this is a buffer region
where we take no data from.  The `field' sample
consists of the stars beyond the outermost concentric circle;
in this work we use this field region to statistically correct the other
two samples of interlopers.
}
\label{fig:schematic}
\end{figure*}

The resultant multi-colour photometry catalogues are
publically available at www.nearbygalaxies.org.
ANGST reaches several magnitudes below the expected
tip of the red giant branch that we will use in Section~4
to determine the distance to Antlia with and is therefore
ideal for our work.  An image of the ACS observations
of Antlia is shown in Fig.~\ref{fig:schematic}.
From ANGST, we use only those stars that have passed
the quality cuts for sharpness and crowding 
(see Dalcanton et al.\ 2009).

We now
divide the data up into three samples (see Fig.~\ref{fig:schematic}): 
an inner sample (covering the
very centre of the galaxy and the highest stellar density regimes; $r<0.02$ deg);
an outer sample (at larger radii from the galaxy centre and lower 
stellar densities; $0.02<r<0.04$ deg); and a `field' sample 
(where we assume the contribution
from the Antlia Dwarf Galaxy is minimized and the majority of the stars
are likely foreground Milky Way stars; $r>0.045$ deg).  
These divisions provide broad analogues of divisions made by
Aparicio et al.\ (1997) and will facilitate comparison in order to
test (e.g.) the presence of a blue stellar core.
Most of the analysis 
in this work will concentrate on the inner sample and we display a schematic of these
divisions in Fig.~\ref{fig:schematic}. We note that we
intentionally build in a small buffer zone between the outer sample 
and the field sample.

%
%

\section{Contamination subtraction}
At bright magnitudes (F814W$<24$), it is likely that of the order 10's 
of Milky Way stars 
contaminate the colour-magnitude diagram for the Antlia Dwarf Galaxy 
(Aparicio et al.\ 1997).  
As one proceeds to fainter magnitudes, the errors in the photometry increase and
we may expect a higher proportion of stars could be interlopers.  In order
to correct this foreground contamination, we follow the statistical method
of Pimbblet et al.\ (2002) to generate artificial star counts
to correct for contamination.  
Although the method was originally developed to
tackle galaxy clusters at modest redshifts, the concept 
underlying the correction technique is identical in the present case
and has been utilized \& emulated extensively in the literature 
(e.g.\ Tanaka et al.\ 2005; Wake et al.\ 2005; Rudnick et al.\ 2009;
Urquhart et al.\ 2010).  
Here, we outline the pertinent details of the procedure.

%
%
\begin{figure*}
\centerline{\psfig{file=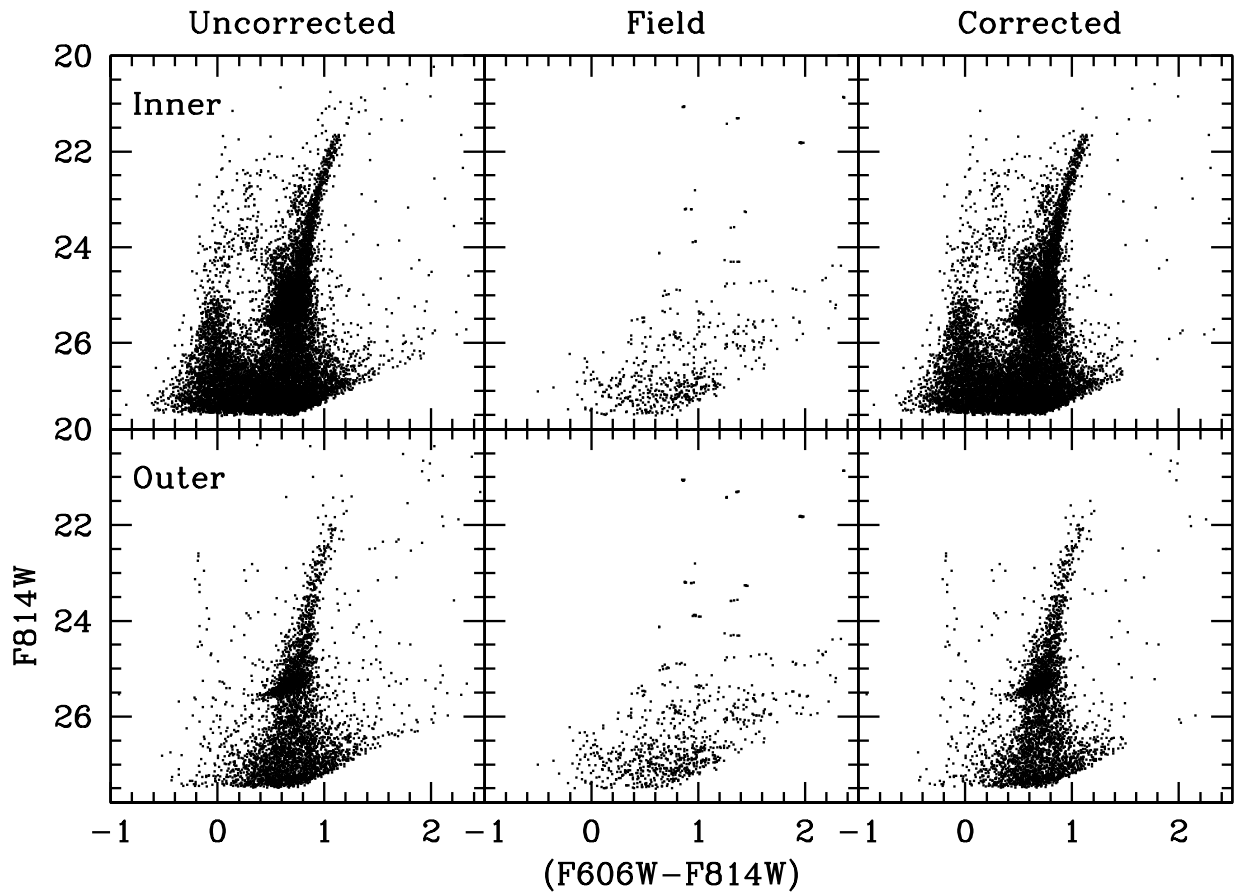,angle=0,width=7in}}
\vspace*{-5cm}
  \caption{Example of the field subtraction technique for the inner
sample (top row) and the outer sample (bottom row).  The original
(uncorrected) colour-magnitude plane is depicted in the left hand column, an
area-scaled field sample is shown in the central column, and
the resultant (corrected)
colour-magnitude plane after statistical correction
is show in the right hand column.  The correction mostly removes
fainter magnitude stars whilst leaving important regions of the plane
(e.g.\ the locus of the tip of the red giant branch) nearly untouched.
}
\label{fig:correction}
\end{figure*}

Each of the three samples are firstly represented on a colour-magnitude grid.
We can then compute the probability of a given object at a particular grid location
being a field object (i.e.\ a contaminant) as
\begin{equation}
P(col,mag)_{Field} = \frac{A \times N(col,mag)_{Field}}{N(col,mag)_{Antlia+Field}}
\end{equation}
where $A$ is an areal scaling factor that matches the size of the field sample
to that of either the inner or outer samples.
In this work, we deviate from the original Pimbblet et al.\ (2002) formulation
by taking care of $A$ through perturbing the field sample.  This is achieved 
adding in extra `field' objects to the field sample until the area covered 
matches that of the inner or outer samples (assuming a fixed stellar density
for the field sample).  We do this by randomly selecting an object from the
field sample and modifying its colour and magnitude by a 
random Gaussian deviation
multiplied by the errors on both colour and magnitude (respectively).  This 
is repeated until the area of the field sample matches that of the inner
or outer samples, as required.
The reason that we choose this approach rather than simply use a factor, $A$,
to scale the probability by is due to the coverage of the field sample 
itself being only modest in size.  
Ideally, we would use an extensive suite of field observations that 
are close to the target (Antlia) so that
gross fluctuations in the stellar foreground do not affect the sample.  
In the present case, this is the only HST data that is available near to 
Antlia\footnote{A search with the high-level science archive 
(hla.stsci.edu) shows that although there are other fields within 10 degrees
of Antlia with the correct combination of filters, none of
them constitute a more appropriate field sample as they are primarily 
targeted on other large, nearby galaxies.}
and we therefore argue that perturbing the field sample in this
manner is more desirable and representative than just applying the scaling.
Once the field sample is area-scaled, $P(col,mag)_{Field}$ is computed 
for each grid in the colour-magnitude plane, using bin sizes of
0.5 in (F606W-F814W) and 1.0 in F814W magnitude.  

$P(col,mag)_{Field}$ gives us the probability of an object at a particular
location on the colour-magnitude plane of being a contaminant.
We use this to determine which objects are members of Antlia 
by generating a random
number and comparing it to $P(col,mag)_{Field}$ for each star.  
If the random number is less than $P(col,mag)_{Field}$, then it
is classed as a contaminant and thrown out.
An illustrative
example of this method is depicted in Fig.~\ref{fig:correction} where we display
the colour-magnitude plane of the original inner and outer samples, 
an area-matched 
field sample and the resultant colour-magnitude diagram after the field
correction.  
As can be seen, relatively few are removed from the critical
region near the tip of the red giant branch, whilst 
$\sim$100's to 1000's 
are removed at both fainter magnitudes and redder colours.
Already we can see that Antlia possesses a significant population
of blue stars and is under-going recent and / or present-day star-formation 
(cf.\ Aparicio et al.\ 1997).  Moreover, the bluer stars are
largely confined to the inner sample whereas the outer sample largely 
lacks this population.
This qualitative observation supports the findings of previous 
works (Aparicio et al.\ 1997; McQuinn et al.\ 2010) and means that they
are not adversely affected by stellar contamination.

The statistical field correction 
is repeated 100 times in a Monte-Carlo fashion 
for both the inner and outer samples to give better statistics for this work.
However, it is apparent from Fig.~\ref{fig:correction} that a large number
of objects are removed in comparison to models of the Milky Way's
stellar distribution (Robin et al.\ 2003).  Part of this
may be intrinsic to the methodology employed and choice of field 
sample. 
By this, we mean that there are likely stars from Antlia contained
in the field sample and will therefore be part of the subtraction.
Ry{\'s} et al.\ (2011) show that red giant stars in the outer halo
of other dwarf galaxies are well fit by 
a de Vaucouleurs profile.  
Using such a profile, at the radii of the field sample
we would expect 0--1 Antlia stars arcmin$^{-2}$ to be present,
which at worst would correspond to $\sim 20$ stars.  Given the
constraints of the data, we are happy to live with this level
of self-contamination.

However, there may be further contamination from other sources (i.e.\ background
galaxies).  Given that galaxy profiles are generally very different from stellar
profiles, we are confident that the ANGST pipeline results in only small
contamination from galaxies for our study at bright magnitudes.
But there could certainly
be an appreciable population of (unresolved) galaxies that masquerade as stars
-- especially at faint magnitudes (cf.\ Radburn-Smith et al.\ 2011).  
To be confused with a star, a
background galaxy would have to have a light profile similar to a
compact dwarf galaxy (cf.\ Gregg et al. 2003; Drinkwater et al.\ 2003).
In the absence of redshifts, it is practically impossible to
differentiate such galaxies from stars.
Alternatively, an unresolved galaxy can be readily confused with a star simply by
being near to the photometric limit.  This type of contaminant would account
for the majority of the low magnitude ``stars'' that are subtracted.
Since we're using a statistical correction, such galaxies should be
properly subtracted assuming that there are no large galaxy clusters
in the background.  

We contend that within the limits of our chosen
method and field sample, the large number of objects removed will have
minimal impact on the parameters of merit (e.g.\ distance)
that we will derive below.

\section{Tip of the Red Giant Branch}
The tip of the red giant branch (TRGB) is an excellent indicator of galaxy 
distance if one can resolve individual stars inside a given target 
galaxy (Lee, Freedman \& Madore 1993; see also Madore \& Freedman 1998 
and references therein).  
Since the tip of the red giant
branch represents the first ascent (core helium flash)
of red giant branch, 
the method is analogous to finding standard candles in
nearby galaxies.  Use of HST imaging combined with this method has
directly lead to accurate determinations of 
distances to nearby systems
ranging from the Large Magellanic Cloud (Romaniello et al.\ 2000)
to NGC~300 (Rizzi et al.\ 2006) and beyond (Dalcanton et al.\ 2009;
Radburn-Smith et al.\ 2011).

One of the strengths of the method is its simplicity: the 
key observable is to determine the 
$I$-band magnitude of the red giant branch
where the luminosity function is abruptly truncated.  
At $I$-band wavelengths, the dependence of this position
along the luminosity function is largely independent 
of metallicity (Da Costa \& Armandroff 1990).  
Further, the TRGB magnitude is only expected to 
vary by $\sim0.1$ mag over a large range of 
of ages (2--15 Gyr; Iben \& Renzini 1983).
Although the TRBG was determined by eye
in its early days (e.g.\ Mould \& Kristian 1986),
a straight-forward
edge detection technique (e.g.\ 
Sakai, Madore \& Freedman 1996; 
see also M{\'e}ndez et al.\ 2002) 
is now more frequently applied to the
data to determine the position of the tip.

There are many approaches to edge detection,
ranging from the Canny (1986) algorithm, Laplacian edge detection,
to more complex methodologies (e.g.\ 
Frayn \& Gilmore 2003; M{\'e}ndez et al.\ 2002).  
In general, the task of
edge detection is a non-trivial endeavour due to
how data is binned and noise properties
present.  In astronomical imaging and two dimensional
edge detection problems, there
are a wide range of edge detection approaches
employed for various ends such as cosmic
ray detection and rejection (e.g.\ 
Laplacian edge detection; see Farage \& Pimbblet 2005;
van Dokkum 2001) and morphological measurements 
to differentiate stars from galaxies (e.g.\ the 
mathematical morphology gradient
operator; see Moore, Pimbblet \& Drinkwater 2006).
Considering the high quality of the colour
magnitude diagrams (Fig.~\ref{fig:correction}), 
the standard Sobel filter approach (Lee et al.\ 1993)
is sound enough for this work (Svalbe, priv.\ comm.).  

\begin{figure}
\centerline{\psfig{file=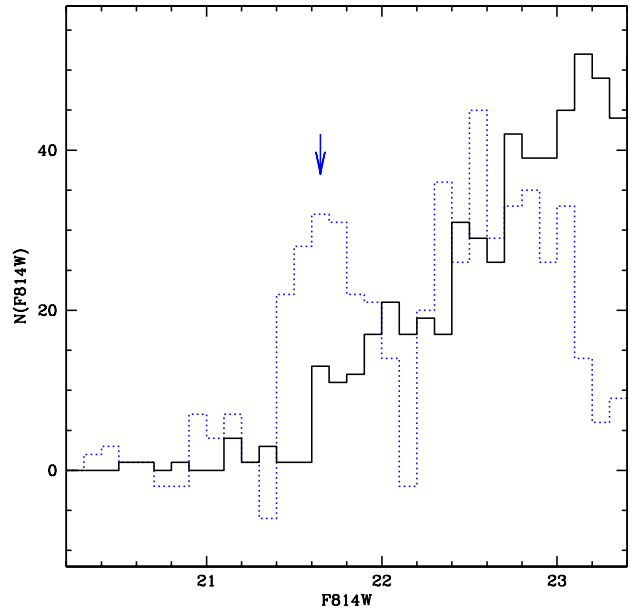,angle=0,width=3.5in}}
  \caption{Example of the response of the Sobel
filter (dotted line) to one of the 
inner sample's realizations of
Antlia's luminosity function (solid line) using
a bin size of 0.1 mag.  The 
downward arrow denotes the peak of the Sobel
filter's response just before the luminosity
function gets truncated which we interpret as the 
location of the TRGB.
}
\label{fig:lf}
\end{figure}

For each of the realizations of the contamination correction
technique (Fig~\ref{fig:correction}), we now compute
the position of the $I$-band (i.e.\ F814W band) 
TRGB by creating a luminosity
function and passing a Sobel filter with a
kernel of [-2,-1,0,+1,+2] across it. 
The luminosity function is limited to $0.6<$(F606W-F814W)$<1.3$
to curtail the influence of non-red-giant stars 
on the resultant TRGB measurement.
This kernel is in keeping with Lee et al.\ (1993; see
also Sakai et al.\ 1996) and will
yield a maximum value for the greatest count discontinuity.
An example of the application of this filtering technique is 
displayed in Fig.~\ref{fig:lf}.

%
%

We now apply this method to all of field subtracted
realizations of the Antlia colour magnitude diagram.
We find a raw (i.e.\ not
extinction corrected) $m_{TRGB} = 21.687 \pm 0.049$ 
for the inner sample, whereas the outer sample
yields $m_{TRGB} = 21.849 \pm 0.010$
where the quoted error is the standard deviation 
from the 100 realizations.
The difference between the inner and outer samples
is likely due to extinction (cf.\ Holwerda 
et al.\ 2009 who report a few tenths mag extinction in 
the outer regions of more massive galaxies).

Using the approach of M{\'e}ndez et al.\ (2002), 
Dalcanton et al.\ (2009) report a raw $m_{TRGB} = 21.642$ --
using the same HST dataset, the only difference being that
they employ the full field of view and avoided
the highly crowded parts of Antlia
-- less than $1\sigma$ away from 
the value that we derive for the inner sample.  
We note that Dalcanton et al.\ (2009) 
also perturbed the stellar data
points by a Guassian random error in Monte Carlo trials
to obtain their result. 
This suggests that the more complex approach of 
M{\'e}ndez et al.\ (2002) who Gaussian smooth 
the luminosity function prior to applying a 
continuous logarithmic edge detection 
(a subtle modification of the Sakai et al.\ 1996 approach)
does not dramatically
improve the computation of $m_{TRGB}$.  This is likely
due to the fact that this bright part of the colour
magnitude diagram is not significantly affected by
noise. Following Dalcanton et al.\ (2009; inparticular
using the extinction and absolute $M_{TRGB}$ value;
from Table~5),
we compute a distance of $1.31 \pm 0.03$ Mpc to Antlia.

From a search of the NASA/IPAC Extragalactic Database (NED),
our results are completely congruous
with Dalcanton et al.\ (2009) who themselves 
derive a distance of $1.29\pm0.02$ Mpc to Antlia and
the earlier studies of Aparicio et al.\ (1997) who report
$1.32\pm0.06$ Mpc,
van den Bergh (1999) at $1.33 \pm 0.10$ Mpc,
Blitz \& Robishaw (2000) with $1.24 \pm 0.07$ Mpc,
and Tully et al.\ (2006) with
1.25 Mpc\footnote{Tully et al.\ (2006) do not quote 
an error on Antlia's distance.  However, they do
quote $I_{TRGB} = 21.60 \pm0.13$, which would produce
an error on their distance
that is bigger than ours by a factor of 2--3.}.
However, it is somewhat smaller than 
the value of $1.51\pm0.07$ Mpc reported by
Piersimoni et al.\ (1999) who obtained
$I_{TRGB} = 21.7 \pm 0.1$ from ground-based
observations; although the amount of 
dust affecting the galaxy may be significant
and cause the distance estimate to vary
by as much as 0.1 Mpc 
(see Sarajedini, Claver, \& Ostheimer 1997
who quote a range of distances from 1.24 to 1.33 Mpc
depending on the amount of dust present).  
Finally,
we note that Whiting, Irwin \& Hau (1997)
found a distance of $1.15\pm0.10$ Mpc which
seems the most discrepant distance in the literature.
We suggest the depth and 
photometric accuracy of HST 
is a prime factor in the difference to 
this earlier analysis.

We note that our distance 
places Antlia over 300 kpc from NGC~3109, thereby
implying little present interaction.

\section{Morphology of Antlia}

Motivated by tying together the structure of a galaxy
to its formation and evolutionary history, 
the use of quantitative morphology has bloomed 
over the past few decades.  
One of the most widely adopted (and straight forward)
approaches to quantitative galaxy morphology
is the use of the model-independant `CAS' paradigm 
(Conselice 2003; see also Conselice 
2006 and references therein).  We also make use
of the non-parametric
Gini ($G$) and $M_{20}$ coefficients
(see Lotz et al.\ 2004) to complement the CAS
paradigm.
Before we compute the CAS values,
we must first turn the contamination corrected
realizations of Antlia stars
in to new images.

\subsection{Image Creation}
\begin{figure}
\centerline{\psfig{file=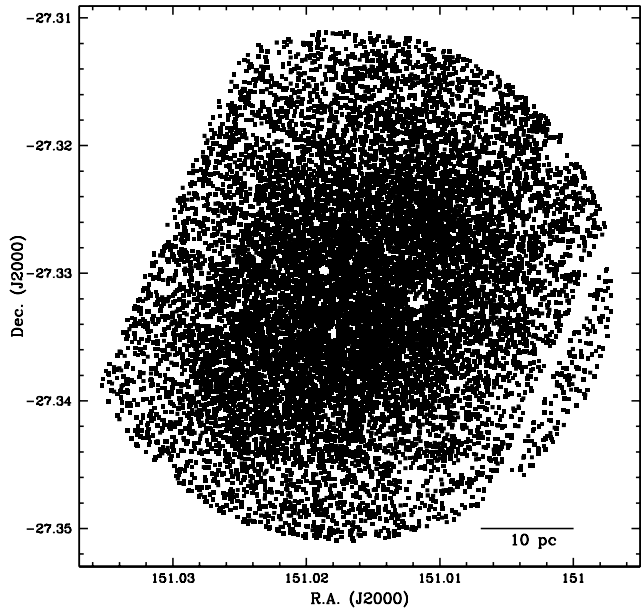,angle=0,width=3.5in}}
  \caption{Spatial distribution of one of the 
contamination subtracted realizations
of the inner sample.  
The scale bar denotes 10 pc at the TRGB distance of Antlia.
Several issues are immediately
apparent if we are to compute CAS parameters for
this distribution, including the gap between chips
and obvious holes in the distribution due to 
saturated stars near the centre of 
Antlia (cf.\ Fig.~\ref{fig:schematic}).
}
\label{fig:image1}
\end{figure}

Each realization of Antlia resulting 
from our correction technique contains both (RA,Dec)
and the HST pixel (x,y) positions.  In principle, 
we can simply bin up these positions to create a 
new image of Antlia for each contamination-corrected 
realization. 
But as Fig.~\ref{fig:image1} demonstrates,
there are several issues to deal with in
binning up the image.

%
%

Firstly, there is a very obvious gap between the 
HST chips, as well as an edge on the opposite
side.  Secondly, there are clear holes in the
stellar distribution due to large, saturated stars
in the original HST image (cf.\ Fig.~\ref{fig:schematic})
Thirdly, the data need to be binned up into larger
pixels, but the exact amount of binning required
is unclear.

\begin{figure*}
\centerline{\psfig{file=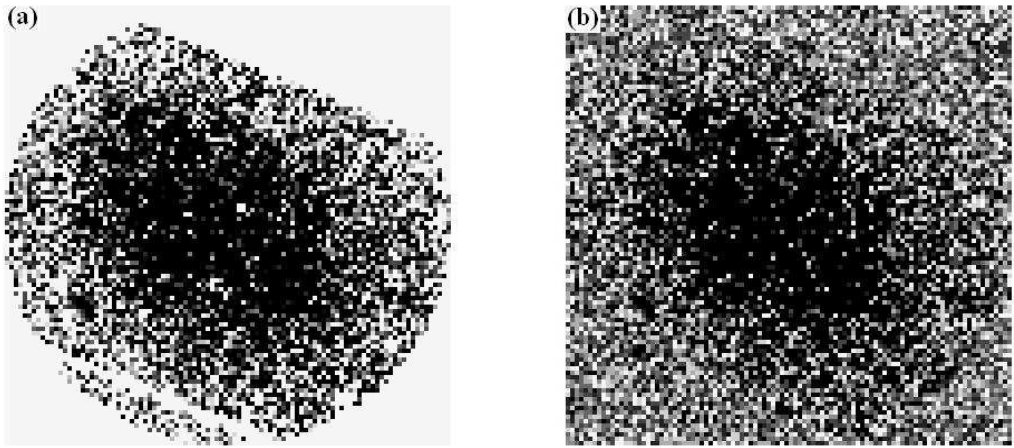,angle=0,width=7.in}}
  \caption{Illustrative examples of binned realizations of one field
corrected inner sample; (a) each pixel is weighted according to 
the brightness of the stars falling within the pixel;
(b) as for (a), but incorporating
random noise at a level comparable
to the outskirts of the galaxy
in pixels whose value would otherwise be
zero and covering over the holes in the pixel distribution
due to saturated stars (Fig.~\ref{fig:image1}).
}
\label{fig:image2}
\end{figure*}

We tackle the latter question first.  When the original
CAS parameters were tested out, they were applied to
and benchmarked against the
Frei et al.\ (1996) dataset (Bershady et al.\ 2000;
Conselice 2003; see also Conselice et al.\ 2000).  
The Frei et al.\ (1996) sample consists of 113 nearby
galaxies imaged in multiple pass-bands from ground
based telescopes, with various auxiliary data
available (Conselice et al.\ 2000).  
Therefore, it makes logical sense to attempt
to bin up our data so that it appears to be
at the same distance as these benchmark galaxies.  
From Fig.~4 and Table~1 of Frei et al.\ (1996),
the mean heliocentric recession velocity of the sample
is a little above $\sim1000$ kms$^{-1}$.  
Given our calculation of the distance to Antlia above,
we would have to move Antlia outward by a factor of
$\sim5$ to match its physical scale to angular size ratio
to that of a typical Frei et al.\ (1996) galaxy.  Rounding
down, this yields $\sim$ $100 \times 100$ pixels
to fit the points displayed in Fig.~\ref{fig:image1} into.
To populate the bins, we weigh each pixel proportional 
to the brightness of   
the stars contained within (Fig.~\ref{fig:image2}).

To ensure that chip edge effects are minimized, we replace
all pixel values that were set at zero with a random value 
scattered about the mean of the pixel values in the edge 
regions.  
This makes the edges of the inner sample blend in with
surrounding noise.
Finally, to take care of the holes in the pixel
distribution due to the saturated stars near the centre
of the stellar distribution, we draw values from nearby
populated pixels to interpolate over then.  An example of 
one of the images this process produces can be seen in
Fig.~\ref{fig:xcont}.  In this image, the 100 pixels
cover $\approx88$ pc at the TRGB distance of Antlia.
This process is performed for both the F814W and F606W bands.
We suggest that this resultant
image is sufficient for the subsequent application of
the CAS quantitative morphology algorithms.

%
%

\subsection{Concentration ($C$)}
The first of the CAS parameters that we measure 
is the concentration of the galaxy, $C$.
Central light concentration can be measured in
a variety of ways (e.g.\ Kent 1985; Abraham et al.\ 1994),
but each provides sensitivity to different galaxy 
morphological populations (Conselice 2003)
and can readily be used for star-galaxy differentiation  
(e.g.\ Pimbblet et al.\ 2001).
We follow Conselice (2003) and Bershady et al.\ (2000)
and define 
\begin{equation}
C = 5~log_{10} (r_{80} / r_{20})
\end{equation}
where $r_{80}$ and $r_{20}$ is 80\% and 20\% of 
the curve of growth radii 
(Bershady et al.\ 2000; see also Petrosian 1976;
Wirth, Koo \& Kron 1994).

For our 100 realizations of Antlia, we find
that $C = 2.003 \pm 0.004$ for F814W 
and $2.000 \pm 0.003$ for F606W (where the quoted
error is the standard deviation of $C$ from the 
100 realizations). 
This is a remarkably low
central concentration, amongst the
lowest values produced through this method
(for comparison, 
elliptical galaxies tend to give $C>4$ whereas
disk dominated galaxies produce values in
the range $3<C<4$ for $R$ band images which sits between
the F814W and F606W bands; Conselice 2003).
Conselice (2003) suggest that this value
may be consistent with irregular galaxies, very late
disk types, as well as dwarf elliptical galaxies.

\subsection{Asymmetry ($A$)}
The second CAS parameter, asymmetry is a very 
straight forward measure of how symmetric the galaxy is.
Formally, we follow Conselice (2003) and define it as
\begin{equation}
A = abs (I-R) / I
\end{equation}
where $I$ is the original image, 
and $R$ is $I$ rotated through 180 degrees about
its centre.  Both $I$ and $R$ around found by
summing over all pixel values in the image.
$A$ has been shown to correlate well 
with both morphological type and colour of a galaxy
(Conselice et al.\ 2000; Conselice 2003) with lower $A$
values denoting both redder colours and morphologically 
earlier types.  We refer the reader to Conselice (2003)
for more detail about this parameter.

There are usually issues with its derivation that are
noteworthy, however.  Firstly, the exact choice of the 
centre of the image can result in dramatic changes to
the value of $A$ -- Conselice et al.\ (2000) report
that even a 1 per cent shift of the centre can cause
$\sim50$ per cent change in $A$.  The usual manner to
cope with this is to find which of the 8 
pixels surrounding the nominal central pixel has
the minimum value of $A$.
In this work, we use the luminosity-weighted centre of 
Antlia quoted by Whiting et al.\ (2007).
This should be accurate to within less than $0.1$'
(i.e.\ sub-pixel accuracy 
in our binned up image) and adequate
for our purposes.

The second main issue with the computation of $A$ is
background subtraction.  Typically one needs to sample
a nearby `blank' region and subtract off the nominal `sky' 
background from the image.  We note that we have already
(partially) performed this operation through use of 
our statistical correction technique.  
The noise we added at the image 
creation step (Fig.~\ref{fig:image2}) is a real issue though, 
since we are normalizing the value of $A$ by $I$ (see above).  
We therefore subtract off the average noise value 
from the denominator of the above equation
in our computation of $A$.

This results in a mean value of $A = 0.0633 \pm 0.0004$ 
for F814W and $0.0479 \pm 0.0004$ for F606W
for all of our realizations.  This value of $A$ is 
consistent with a morphological classification of an
early-disk type ($\sim$Sa) according to the figures
presented in Conselice et al.\ (2000; in turn based on the
Frei et al.\ 1996 sample).
In colour space, Antlia is most consistent with 
an integrated ($B-V$) colour of $\approx 0.8$ 
(see equation 5 of Conselice 2003) -- very much 
inline with other early-disk types.

\subsection{Clumpiness ($S$)}
The clumpiness of a galaxy physically traces patchiness
of the light distribution of a galaxy at high spatial
frequencies.  Given that star-formation occurs in
clumps and clusters that later disperse 
(Harris et al.\ 2001), Conselice (2003) shows that $S$
correlates well with (recent) star-formation.
Formally, 
\begin{equation}
S = (I - B) / I
\end{equation}
where $I$ is the initial image and $B$ is a blurred version
of $I$.  Hence for smooth, elliptical galaxies, $S$ should
take values of $\sim 0$.
The $B$ image is produced by smoothing $I$ with a tophat 
filter of width $\sigma$.  The exact value of $\sigma$ can,
in principle, take on any value in order to better
probe clumpiness on a variety of scales.  
Here, we follow Conselice (2003) and 
Bershady et al.\ (2000) and set $\sigma$
to $0.3~r(\eta=0.2)$ --
the equivalent of $\sigma \approx 15$ pixels\footnote{Following
Conselice (2003), we adopt 
Petrosion's (1976) concept of deducing the rate of change 
of the enclosed light as a function of radius.  As with Conselice (2003),
we use the
inverted form, $\eta(r)=I(r)/<I(r)>$.  Hence the radius 
$r(\eta=0.5)$ would be interpreted as approximately 
the half-light radius, $r_e$.
}.
As with $A$, we need to subtract off the (known)
average value of the noise that we
added to the constructed image in order to compute $S$.

We find that $S = 0.0769 \pm 0.0008$ for 
F814W and $0.0580 \pm 0.0006$ for F606W in
our 100
realizations of the inner sample's contamination 
corrected colour-magnitude diagram.  
This value of $S$ indicates an early-disk type morphology
(i.e.\ Sa), and therefore agrees with the interpretation
of $A$, above.

\subsection{Gini ($G$)}
When applied to astronomical imaging, the 
Gini coefficient, $G$, measures how strongly nucleated
(or, conversely, how patulous) a distribution of pixels is
(Gini 1912; Glasser 1962; Abraham et al.\ 2003; Lotz et al.\ 2004; 
Law et al.\ 2007).  Following Glasser (1962), the Gini
coefficient is defined as:

\begin{equation}
G = \frac{1}{\bar{X}N(N-1)} \sum_{i=1}^{N} (2i-N-1) X_i
\end{equation}
where the pixel values, $X_i$, are sorted from smallest
to largest before summing over all $N$ pixels.  A value of
$G=1$ is therefore interpreted as a single pixel possessing
the entire flux of the image whereas at the opposite extreme,
$G=0$, each pixel has an equal share of the flux.

We find that $G=0.3893 \pm 0.0005$ in the F814W band 
and $G= 0.3937 \pm 0.0006$ in F606W (again, where
the error is the standard deviation on our 100
realizations of Antlia).  This is a very low value
of $G$ (cf.\ Abraham et al.\ 2003). 
We note here that
$G$ is sensitive to signal-to-noise ratios 
and dependant on the aperture used (Lisker 2008).
However, our derived value is very much
in-line with the interpretation of $C$ (above) of
Antlia being a very late type.

\subsection{$M_{20}$}
The $M_{20}$ parameter measures the second order moment
of the brightest 20 per cent of the flux of the image
(Lotz et al.\ 2004) and is somewhat 
more sensitive than $C$ to signatures of 
mergers such as multiple nuclei (see 
F{\"o}rster Schreiber et al.\ 2011 for a recent example
of its application).  The total second order moment
of the pixels is defined as:
\begin{equation}
M_{total} = \sum_{i}^{n} M_i = \sum_{i}^{n} f_i [ (x_i-x_c)^2 + (y_i-y_c)^2  ]
\end{equation}
where $f_i$ is the flux (pixel value) of each pixel and
the subscript $c$ denotes the central pixel, as defined
for $A$ above.  In order to obtain $M_{20}$ the pixels are 
rank-ordered (brightest first) 
and $M_i$ summed until 20 per cent of the total
pixel values is reached, thus:
\begin{equation}
M_{20} = log_{10} (\sum_i M_i / M_{total})
\end{equation}
whilst $\sum_i f_i<0.2 f_{total}$. For the F814W band,
we find $M_{20} = -1.168 \pm 0.004$, whilst
$M_{20} = -1.207 \pm 0.007$ in the F606W band.  These are
modest values for $M_{20}$ and suggestive of late-type
disks (e.g., see Lotz et al.\ 2004; 2006; 2008), although we
explicitly 
note that these previous studies investigating $M_{20}$ do not
include dwarf elliptical populations.  However, in general $G$ and
$M_{20}$ tend to anti-correlate and our derived values 
reflect this general trend for `normal' undisturbed galaxies.

\subsection{Surface Brightness}
In the above analysis, we have applied the CAS
formalism to HST imaging and
compared it to parameters presented by Conselice (2003)
that are based on the Frei et al.\ (1996) dataset.
But these two datasets are not directly comparable.
For the HST imaging of Antlia, Sharina et al.\ (2008)
demonstrate that the limiting surface brightness
within the central arcmin
of the Antlia HST/ACS dataset is 
$\mu_V = 25.3$ mag arcsec$^{-2}$ and
$\mu_I = 25.0$ mag arcsec$^{-2}$.  Further, the 
peak surface brightness is given as
$\mu_V = 23.9$ mag arcsec$^{-2}$ and
$\mu_I = 23.3$ mag arcsec$^{-2}$ (Sharina et al.\ 2008;
their Table~1).

However, for the Frei et al.\ (1996) dataset,
Bershady et al.\ (2000; their Table~3) show that the 
average surface brightness varies from $mu_B = 20.3$ 
to $mu_B = 21.7$ mag arcsec$^{-2}$ within one half-light
radius for morphologically elliptical to late sprial and irregular
(respectively). Therefore, we are 
unable to compare our results
in a direct manner to the (bright) Frei et al.\ (1996) sample
through cutting our Antlia images to the same surface 
brightness limits.

Conselice (2003; inparticular Table~3) 
extend the Frei et al.\ (1996) sample 
with supplemental dwarf elliptical galaxies from 
Conselice, Gallagher \& Wyse (2003).  For this 
modest dataset, the surface brightness limits 
\emph{are} comparable since Conselice et al.\ (2003)
by design only select dwarfs with $\mu_B > 24.0$
mag arcsec$^{-2}$.

\subsection{Interpretation}
Before we interpret the results, we note an important caveat. 
In the above application of CAS, $G$ and $M_{20}$ to our dataset, we
have not strictly followed the literature prescription
since we have not segmented our image before calculation.  This
leads to more noise in our data than our 
comparison sample (Conselice 2003; see also Lotz et al.\ 2004; Lisker 2008).
Performing a segmentation operation on our images has multiple
issues since (a) each realization arises from a different contamination 
correction; and (b) the images are already (artificially) truncated
to the inner regions of Antlia (Fig.~\ref{fig:schematic}).
To evaluate if segmenting the image would make any difference
to our results and interpretations, we follow Hambleton et al.\ (2011)
and re-compute the asymmetry as:
\begin{equation}
A = \frac{\sum I-R}{\sum I} - \frac{\sum B - B_R}{\sum I}
\end{equation}
where $B$ is the `background' of the image, taken to be a
$10 \times 10$ pixel area in a corner of the image, and $B_R$
is $B$ rotated through 180 degrees.  We find that the change
in $A$ due to subtracting off the background term is  
$<<1$ per cent -- the dominant factor in the variation
of $A$ is found to come from the contamination correction.  Similar 
changes are found for $S$ and $G$ using an analogous approach.

The combination of CAS with $G$ and $M_{20}$ allow
us to investigate how disturbed Antlia is and therefore
assess any degree of recent morphological change (e.g., due
to NGC~3109). 
Conselice (2003) derives the relationship
between $A$ and $S$ for the Frei et al.\ sample
as $A=(0.35\pm0.03) \times S + (0.02\pm0.01)$.  Hence 
given our F814W value of $S=0.0769$, the predicted value 
of $A$ would be $0.0469$, 0.0164 (i.e.\
less than $2\sigma$) from our
derived value.  Large deviations away from the predicted
value of $A$ would be suggestive of galaxies which
are involved in mergers (Conselice 2003).  
We are also able to use $G$ and $M_{20}$ to evaluate
if there are any signatures of recent major merging 
activity (modulo the caveats given by Lisker 2008).  
For instance, Lotz et al.\ (2008) suggest that
galaxies at higher redshifts with $G>-0.14 M_{20} +0.33$ are
mergers -- Antlia lies far away from this regime.  
Comparison to Lotz et al.\ (2004; inparticular their Fig.~9) 
underscores that Antlia is a relatively `normal' undisturbed galaxy.

There are further ways in which mergers (at least of more massive
galaxies) have been quantified using combinations of these parameters.
For example, Lotz et al.\ (2004) define ULIRG mergers as 
$G>-0.115 \times M_{20} +0.384$ and 
$G>-0.4 \times A +0.66$ (or $A>0.4$; cf.\ Conselice 2003 who
use $A>0.38$). None of these criteria are met for our
analysis of Antlia.
One final way in which morphological disturbance can
be quantified from these parameters
is introduced by Holwerda et al.\ (2011a) from their
quantitative analysis of H{\sc i} morphologies: 
the Gini coefficient for the distribution of second
order moments, $G_M$.  This is defined in a completely analogous way
to the original Gini coefficient, but replacing
pixel values with $M_i$ values:
\begin{equation}
G_M = \frac{1}{\bar{M}N(N-1)} \sum_{i=1}^{N} (2i-N-1) M_i
\end{equation}
With the caveat that this parameter has not been fully 
explored for an optical
dataset (e.g.\ Frei et al.\ 1996) yet, we find that
for all the realizations of our background correction $G_M<<0.6$.
This indicates a lack of interaction using Holwerda et al.'s (2011a)
definition.

While the Antlia Dwarf Galaxy may be may be
star-forming to a degree, we suggest that 
it does not show clear morphological
signatures of interaction under \emph{any} combination 
of the quantitative morphological parameters investigated here.
Yet that does not preclude historic interaction.
A number of studies have been made describing how long
a timescale the parameters investigated in this work
may show signatures of interactions after a merger event
(Holwerda et al.\ 2011b;
Lotz et al.\ 2010a; Lotz et al.\ 2010b; Conselice 2009).  
But in general, none of them investigate dwarf galaxies.
Therefore although it is likely to have been at least $>\sim$0.5 Gyr
since Antlia's last significant interaction with another galaxy
based on these studies, further simulations are urgently needed
in to how long such signatures remain observable 
for dwarf galaxies.

Overall, we suggest that 
our analysis is contributing evidence that NCG~3109 is
having little effect on Antlia and may not be its
satellite (as opposed to
van den Bergh 1999; see also Lee et al.\ 2003;
Barnes \& de Blok 2001).

\begin{figure}
\centerline{\psfig{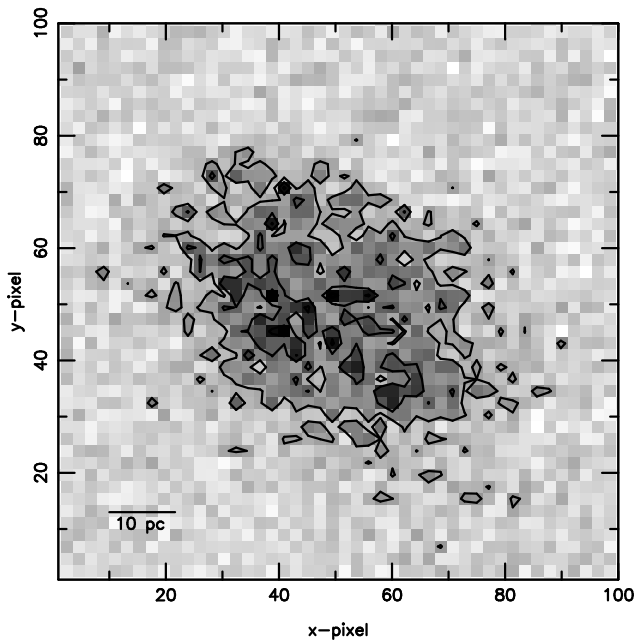}}
  \caption{Contours of smoothed
stellar luminosity density in one realization
of the Antlia Dwarf Galaxy (the contours are set
at arbitrary levels for illustrative purposes).
The side of the box is $\approx 0.04$ deg, or
$\approx88$ pc at the TRGB distance of Antlia.
Antlia appears to have
multiple peaks in stellar density
throughout which account for the low value of $C$.
When combined with the numerical values of $A$, $S$, and $G$,
Antlia is unambiguously determined to be of dwarf elliptical
morphology under the CAS paradigm.
}
\label{fig:xcont}
\end{figure}

We now turn to Antlia's overall morphology.
The values derived for both $A$ and $S$ indicate
a galaxy with an early disk -- approximately
an Sa morphology.
When we combine the result for concentration, $C$, and the Gini
coefficient, $G$,
to the other two parameters, we are able to
unambiguously resolve the morphology
of Antlia in to the dwarf elliptical category (Fig.~15
of Conselice 2003).  The value of $C$ is much too small
to be that of an early-type disk (expectation: $C=3.9\pm0.5$;
Conselice 2003).
Indeed, Conselice (2003) 
gives expected values for dwarf ellipticals as 
$C=2.5\pm0.3$;
$A=0.02\pm0.03$;
$S=0.00\pm0.06$ based on their extended sample of 
cluster dwarf ellipticals observations
(Conselice, Gallagher \& Wyse 2003).
The reason for the
low value of $C$ (and indeed, $G$) becomes obvious from an inspection
of a smoothed contour plot of one of our
realizations (Fig.~\ref{fig:xcont}) which
displays multiple (yet small) peaks in stellar density. 
These peaks also explain the modest value of $M_{20}$.

The developing picture of Antlia is that it has been
relatively isolated in space for some time.  
We confirm that it has a modestly blue colour 
and stellar population coupled with 
an appreciable on-going star-formation rate 
at its core
(Fig.~\ref{fig:correction}), 
agreeing with, e.g., Aparicio et al.\ (1997)
and McQuinn et al.\ (2010).
Although tidal interactions can in
principle transform a dwarf irregular
galaxy in to a dwarf spheroidal and help
promote their star-formation rates, 
it would take several Gyr to do so
(Pasetto, Chiosi \& Carraro 2003).
Whilst it may be the case that Antlia has
undergone historic interactions to
alter its morphology, they are likely
to have been several Gyr ago at minimum.
Indeed, Weisz et al.\ (2011) report that
the typical dwarf galaxy in the local
group has formed the bulk of their stars
by $z\sim2$ and that the differences seen
in star-formation histories between dwarfs
become most pronounced during the past Gyr.
A search for tidal debris in the vincinity of
both Antlia and NGC~3109 would appear a
prudent next step to independently 
confirm the lack of recent interaction.
%
%

\section{Conclusions}
This paper has successfully 
transferred a statistical correction
technique from galaxy cluster studies (Pimbblet et al.\ 2002)
and applied it to the Antlia dwarf galaxy -- a recently
discovered member of the Local Group -- in order
to make new determinations of its distance, 
morphology and formation history. 
Our main findings are:

(1) Using the tip
of the red giant branch standard candle method,
we compute a new distance to Antlia of $1.31 \pm 0.03$ Mpc
that places it at the outermost location of the Local Group.
This value is
in excellent agreement with earlier works (Dalcanton et al.\ 2009;
Aparicio et al.\ 1997), even though we have used a
simple Sobel filter approach to TRGB edge detection.
This distance places Antlia over 300 kpc from NGC~3109.

(2) The colour-magnitude diagram of Antlia qualitatively shows that
the galaxy is presently forming stars (or has been within
the past few 10's of Myr).  The bluer stars are concentrated
in the inner regions, whilst the outer regions ($r>0.02$ deg)
are largely devoid of such blue stars.  This agrees with more
detailed studies (e.g.\ Aparicio et al.\ 1997; McQuinn et al.\ 2010)
and means that these blue stars are not contaminants.

(3) We have applied the CAS formalism to a Local Group
member for the first time (to the best of our knowledge)
and complemented these parameters with $G$ and $M_{20}$
to better derive its morphology
in a quantitative manner and evaluate its formation
history.  We find $C=2.0$, $A=0.063$, $S=0.077$,
$G=0.39$, and $M_{20}=-1.17$
in the F814W band 
(with errors less than $0.1$ per cent across the
100 realizations of the background subtraction)
for Antlia which places it into the category 
of a classic dwarf elliptical galaxy.

(4) Antlia has probably
not had recent merger events with other
galaxies as evidenced by the insignificant deviation
of its asymmetry ($A$) away from the predicted value
based on its clumpiness ($S$).  This is underscored 
with a complementary analysis of $G$ and $M_{20}$ in concert 
with one-another which similarly indicates a lack
of recent interaction.  We tentatively suggest 
that this is contributing evidence against Antlia 
being the satellite of NCG~3109, particularly when 
combined with the new distance derived for Antlia which
means it is over 300 kpc away from NCG~3109.
This does not necessarily preclude a previous interaction,
however.

At the very edge of the Local Group, Antlia 
appears to have been relatively un-touched by 
recent tidal interactions.  Despite having quantitative
morpological values of a classic dwarf elliptical, it
is likely under-going star-formation at the present day
or in the recent past ($\sim 10$'s of Myr; 
cf.\ Aparacio et al.\ 1999; McQuinn et al.\ 2010).  

This work is very much a pilot study.  
It is our intent to perform analogous analyses on 
more ANGST galaxies in a similar manner that would yield
(e.g.) a derivation of new distances to 
other Local Group members.

\section*{Acknowledgements}
We sincerely thank the referee, Benne Holwerda, for
the thorough feedback that has improved earlier versions
of this manuscript.
Some of the underlying ideas in this manuscript
(notably the developed background correction technique) 
have been progressively refined for the best part
of a decade.  We would like to thank various people
who have influenced our thinking over that time, including
our LARCS collaborators, 
Tadayuki Kodama, Michael Bulmer, Richard Bower, Jason Moore, 
Heath Jones,
and Chris Conselice (amongst many others).
We also thank Imants Svalbe for additional insight about the 
edge detection algorithms discussed in this work.

Based on observations made with the NASA/ESA Hubble 
Space Telescope, obtained from the data archive at the 
Space Telescope Science Institute. STScI is operated by 
the Association of Universities for Research in Astronomy, 
Inc.\ under NASA contract NAS 5-26555.

This research has made use of the NASA/IPAC Extragalactic 
Database (NED) which is operated by the Jet Propulsion 
Laboratory, California Institute of Technology, under 
contract with the National Aeronautics and Space Administration.

Finally, and most importantly, we are very grateful to the 
ANGST team for their work in creating the excellent legacy 
data products that this work has used and benefitted from.

\end{document}